\documentclass[aps,prl,twocolumn,superscriptaddress,showpacs]{revtex4-1}

\usepackage{longtable}
\usepackage{graphicx}
\usepackage{subfigure}
\usepackage{dcolumn}
\usepackage{bm}
\usepackage{amsmath}
\usepackage[psamsfonts]{amssymb}
\usepackage{lineno}
\usepackage{color}

\graphicspath{{./}{./Graphs/}}

\newcommand{\etal}{\emph{et al.}~}
\newcommand{\rhounits}{${\mu\Omega}cm$~}

\begin{document}

\title{Kondo-like behavior near the metal-to-insulator transition of nano-scale granular aluminum}

\author{N. Bachar}
\email[]{nimrodb7@post.tau.ac.il}
\affiliation{Raymond and Beverly Sackler School of Physics and Astronomy, Tel Aviv University, Tel Aviv, 69978, Israel}

\author{S. Lerer}
\affiliation{Raymond and Beverly Sackler School of Physics and Astronomy, Tel Aviv University, Tel Aviv, 69978, Israel}

\author{S. Hacohen-Gourgy}
\affiliation{Raymond and Beverly Sackler School of Physics and Astronomy, Tel Aviv University, Tel Aviv, 69978, Israel}

\author{B. Almog}
\affiliation{Raymond and Beverly Sackler School of Physics and Astronomy, Tel Aviv University, Tel Aviv, 69978, Israel}


\author{G. Deutscher}
\affiliation{Raymond and Beverly Sackler School of Physics and Astronomy, Tel Aviv University, Tel Aviv, 69978, Israel}

\date{\today}

\begin{abstract}
We show that the normal state transport properties of nano-scale granular Aluminum films, near the metal to insulator transition, present striking similarities with those of Kondo systems. Those include a negative magneto-resistance, a minimum of resistance $R$ at a temperature ${{T}_{m}}$ in metallic films, a logarithmic rise at low temperatures and a negative curvature of $R\left( T \right)$ at high temperatures. These normal state properties are interpreted in terms of spin-flip scattering of conduction electrons by local magnetic moments, possibly located at the metal/oxide interfaces. Their co-existence with the enhanced superconductivity seen in these films is discussed.
\end{abstract}



\maketitle

Granular Al films have been known for many years to have an enhanced superconducting critical temperature. In this paper we show that in such films, conduction electrons interact with localized magnetic moments. This new finding is surprising since coexistence of an enhanced superconductivity with magnetic moments is unexpected.

\par

We present new transport measurements on aluminum films consisting of nano-scale Al grains, about 2~nm in size, weakly coupled through thin Al oxide barriers~\cite{Deutscher1973a,*Deutscher1972,*Deutscher1973}. We find that near the metal to insulator transition (MIT) their magneto-resistance is increasingly negative and scales with (H/T), with an exponent close to 2, up to about 100~K. Additionally, samples having a positive resistance temperature coefficient (metallic behavior) present a minimum of resistance at a temperature $T_{m}$ of several 10~K depending on the film's resistivity and a temperature dependence of the resistance compatible with a logarithmic increase below $T_{m}$. This logarithmic increase is more clearly seen in films whose resistance increases continuously with decreasing temperature. All metallic films near the MIT display a negative curvature of the R(T) curves. These transport properties point out to spin scattering of conducting electrons, as occurs in Kondo systems ~\cite{Monod1967,Beal-Monod1968}. We discuss possible origins of localized magnetic moments and the compatibility of spin scattering of conduction electrons with the enhanced superconductivity seen in these films.

\par

Samples were prepared by thermal evaporation of 99.999\% pure Al pellets from ceramic crucibles under a reduced pressure of oxygen in the range of $1\div3.5\times10^{-5}$ Torr. Substrates of $Si-Si_{2}O$ were cooled by liquid nitrogen during evaporation. The normal state resistivity, $\rho_{RT}$, of the films was controlled by the oxygen pressure used during evaporation and by the evaporation rate. Fine tuning of this pressure allowed a detailed study of the immediate vicinity of the MIT. Samples whose Kondo-like properties are as mentioned above have normal state resistivities ranging from about 100~${\mu\Omega}cm$ up to several 1000~${\mu\Omega}cm$. In that range the grains size does not vary much and is of about 2~nm~\cite{Deutscher1973a,*Deutscher1972,*Deutscher1973}. The films, about 100~nm thick, are three dimensional in the sense that their thickness is more than one order of magnitude larger than the grain size. They are superconducting with a critical temperature of about 3.2~K with a sharp transition (width of about 0.01~K) indicating a high degree of homogeneity.

\begin{center}
    \begin{table}
        {\small
        \hfill{}
        \begin{tabular}{|l|l|c|c|c|c|c|c|c|}
        \hline
        Sample & $\rho_{RT}$ & $\rho_{4.2K}/\rho_{RT}$ & $T_{c}$ & $T_{M}$ & $T_{m}$ & $\Delta \rho$\\
        & $(\mu \Omega~cm)$ & & $K$ & $K$ & $K$ & $(\mu \Omega~cm)$ \\
        \hline
        \hline
        65 & 65.3 & 0.91 & 2.32 & -- & -- & -0.018 \\ 
        130 & 130 & 0.978 & 3.12 & 9.1 & 25 & -0.04 \\ 
        145 & 145.7 & 0.98 & 3.18 & 8.9 & 25.1 & -0.08 \\ 
        202 & 202.3 & 0.981 & 3.05 & 9 & 28 & -0.12 \\ 
        237 & 237.3 & 0.992 & 3.11 & 8.9 & 44 & $*$ \\ 
        310 & 309.5 & 0.998 & 3.16 & 8.3 & 58 & -0.2 \\ 
        323 & 323.1 & 1.004 & 3.15 & 8.2 & -- & -0.25 \\ 
        408 & 408.5 & 1.01 & 3.1 & 7.5 & -- & -0.25 \\ 
        529 & 529 & 1.013 & 3.06 & 9.2 & -- & -0.63 \\ 
        2425 & 2425 & 1.21 & 2.76 & 5 & -- & -2.71 \\ 
        3470 & 3470 & 1.3 & 2.2 & 5.5 & -- & -12.46 \\ 
        \hline
        \end{tabular}}
        \hfill{}
        \caption{Characteristics of selected samples. $\Delta\rho=\rho(14T)-\rho(0T)$ was obtained at 20~K. (*) was not measured for this sample.}
        \label{tb:Tab1}
    \end{table}
\end{center}

\begin{figure}
    \centering
    \begin{minipage}[t]{\linewidth}
        \centering
        \includegraphics[width=\linewidth]{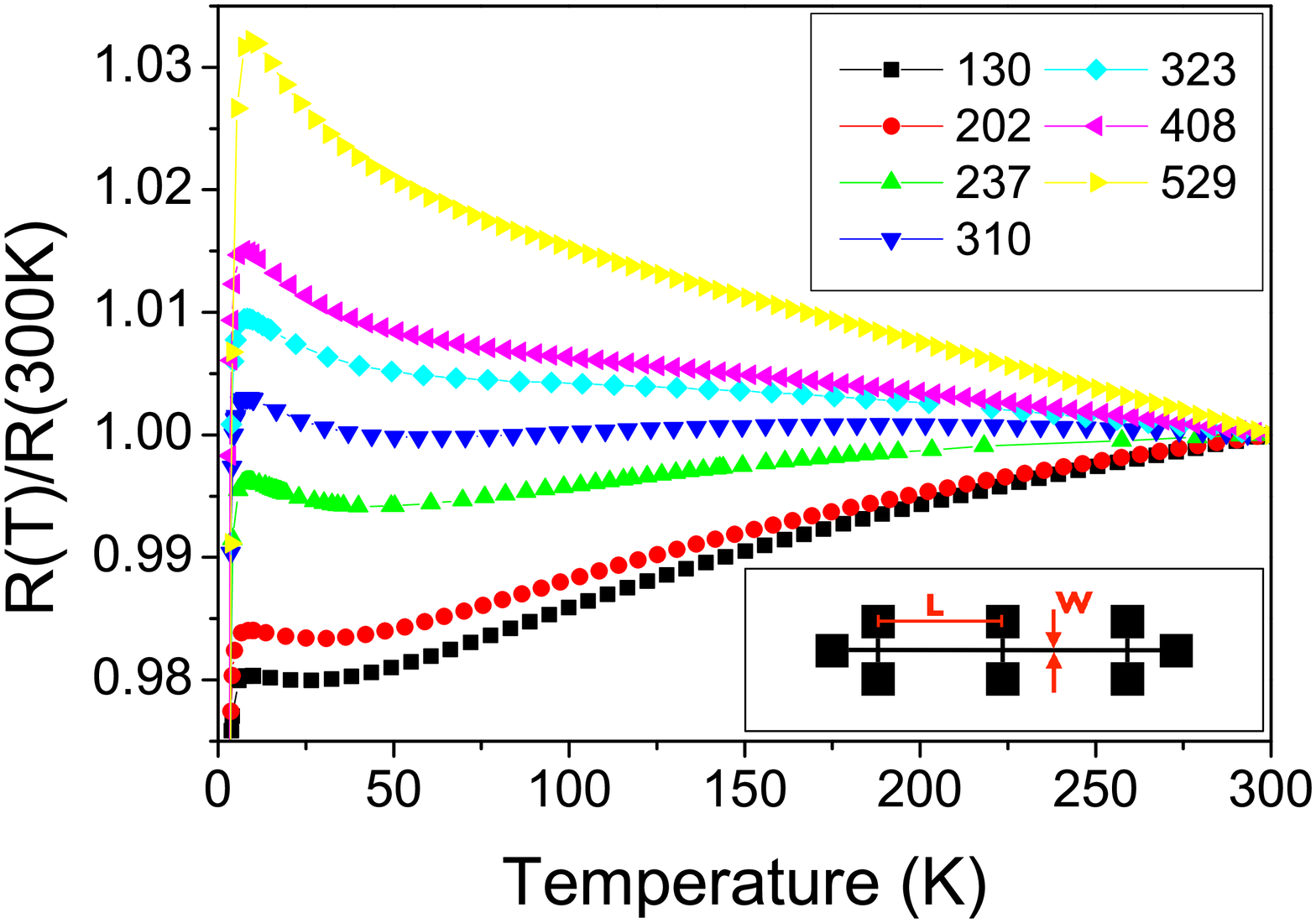}
        \caption{Temperature dependence of the normalized resistivity of selected samples near the metal to insulator transition. A negative curvature and resistivity minimum are observed in films with $\rho \lesssim 300$\rhounits. Inset shows a typical hall bar geometry of our samples with distance between voltage pads of $L=400{\mu}m$ and bar width of $W=10{\mu}m$.}
        \label{fig:Fig1}
    \end{minipage}
    \begin{minipage}[c]{\linewidth}
        \centering
        \includegraphics[width=\linewidth]{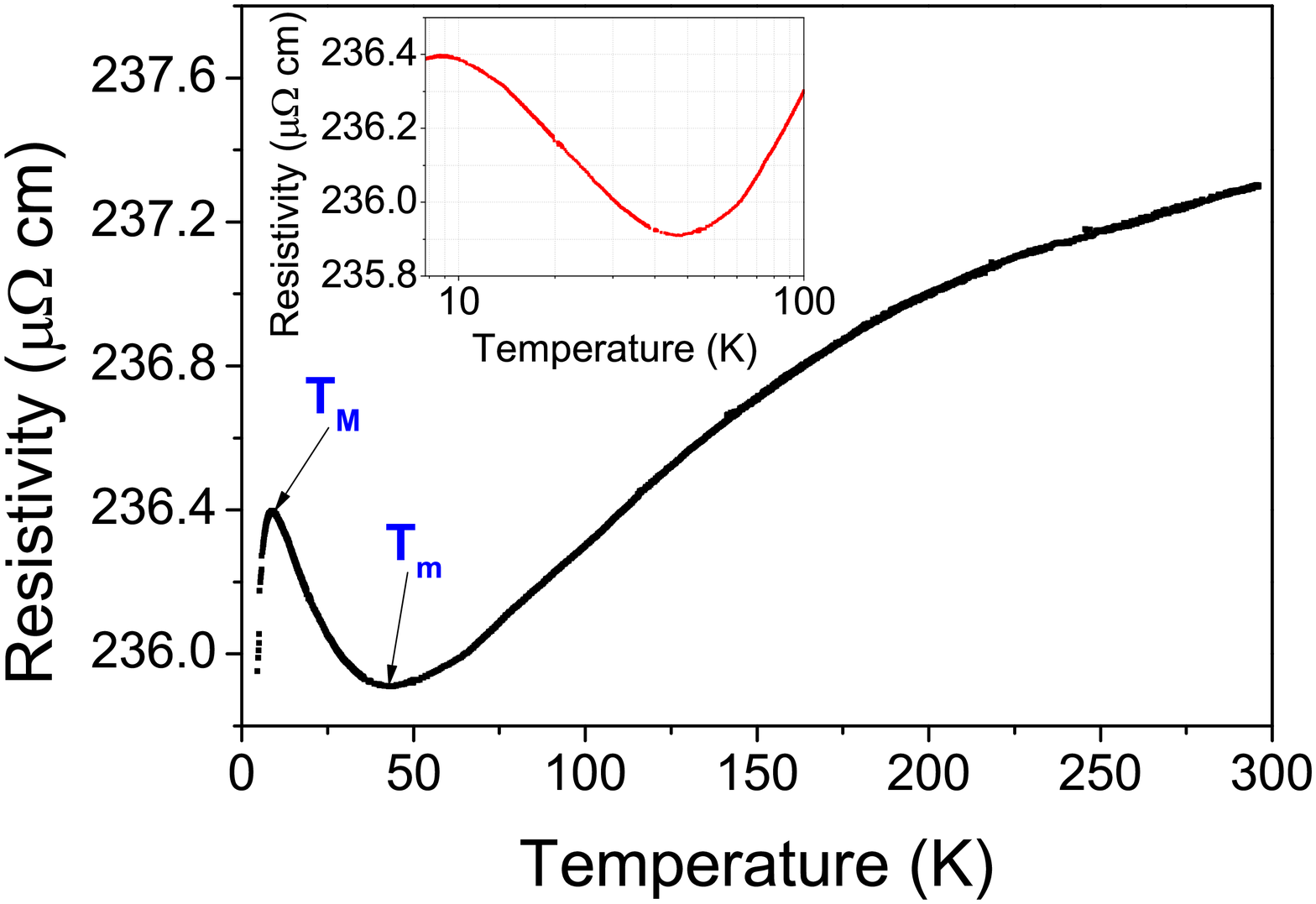}
        \caption{Resistivity as a function of temperature of a metal-like sample (\#237) close to the MIT which shows a negative curvature at high temperature, a resistivity minimum and a log(T) dependence at low temperature.}
        \label{fig:Fig2}
    \end{minipage}
    \begin{minipage}[b]{\linewidth}
        \centering
        \includegraphics[width=\linewidth]{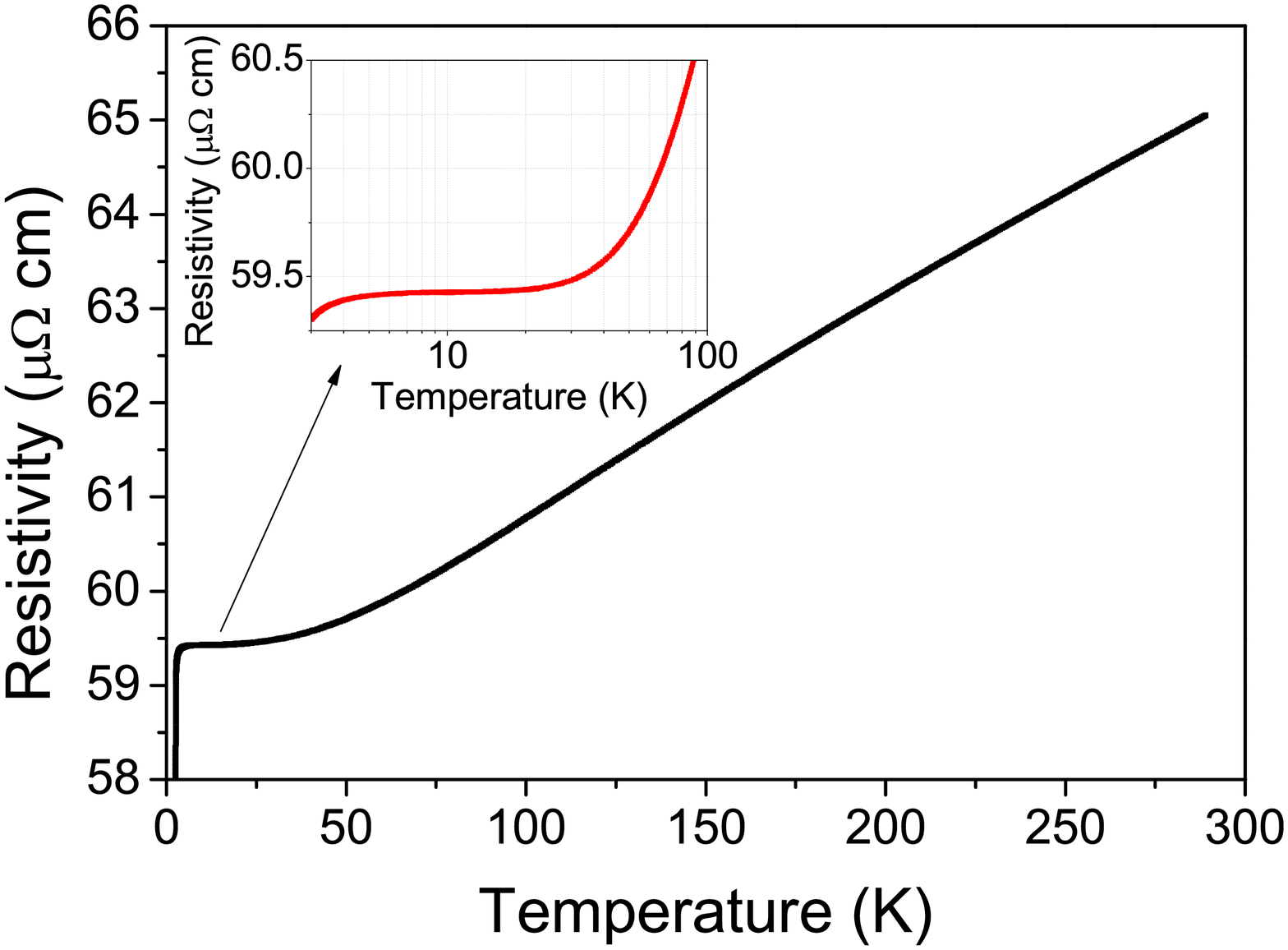}
        \caption{Resistivity as a function of temperature of a low resistivity sample (\#65) far from the MIT. Inset shows the resistivity as a function of log(T). There is no upturn in the resistivity below 20~K.}
        \label{fig:Fig2b}
    \end{minipage}
\end{figure}

Figure~\ref{fig:Fig1} shows the temperature dependence of the resistance of a number of films having a small resistance coefficient of temperature. Table~\ref{tb:Tab1} summarizes some of their properties.

\par

While from earlier measurements~\cite{Cohen1968a} films were simply classified as metallic ($dR/dT > 0$) or insulating ($dR/dT < 0$), a close examination of Fig.~\ref{fig:Fig1} shows that R(T) curves have a non-trivial structure. For resistivities less than $\approx300$~${\mu\Omega}cm$, the behavior is indeed metallic-like, but with a minimum of resistance at a temperature $T_{m}$ that increases with the normal state resistivity. Below $T_{m}$ the resistance reaches a maximum at a temperature $T_{M}$, below which it starts to decrease towards the superconducting transition. We show in Fig.~\ref{fig:Fig2} a more detailed view of the behavior of one of these metallic films. As can be seen in the inset, below $T_{m}$ the temperature dependence of resistance is compatible with a logarithmic increase. Samples having resistivities smaller than $\approx100$~${\mu\Omega}cm$ do not show this low temperature rise (Fig.~\ref{fig:Fig2b}).

For resistivities larger than$\approx300$~${\mu\Omega}cm$, the films resistance rises monotonically as the temperature is reduced. A logarithmic increase of resistance of high resistivity films at low temperatures is observed in these films over a broader temperature range than in metallic films, consistent with earlier findings in similar films~\cite{Deutscher1980}.

\par

Above $T_{m}$ and close to the MIT, R(T) displays a negative curvature at high temperatures on the order of 200~K. This behavior, which is reported here for the first time in granular aluminum, is observed both in metallic-like (Fig.~\ref{fig:Fig2}) and insulator-like (e.g. sample \#408 in Fig.~\ref{fig:Fig1}) films.

\par

All films listed in Table~\ref{tb:Tab1} have a negative magneto-resistance (MR) above $T_{M}$, up to a temperature of the order of 100~K. This negative MR does not saturate up to the highest field reached (14~T in most cases). Below $T_{M}$ the behavior of the MR is more complex, being clearly influenced by superconducting fluctuations that give rise to Ghost Critical Field effect~\cite{Kapitulnik1985}.

\begin{figure}
    \centering
        \begin{minipage}[t]{\linewidth}
        	\centering
            \subfigure[]{\label{fig:fig3a}\includegraphics[width=\linewidth]{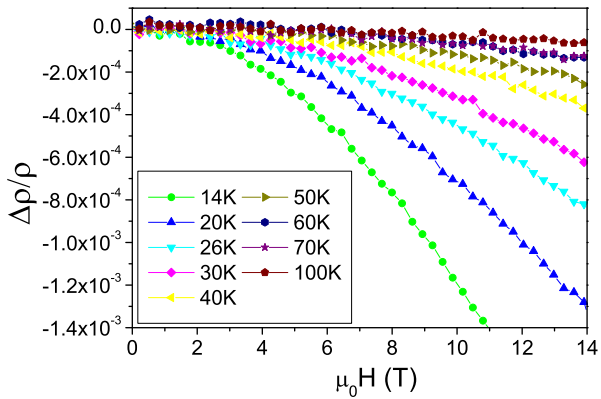}}
        \end{minipage}
        \begin{minipage}[b]{\linewidth}
        	\centering
            \subfigure[]{\label{fig:fig3b}\includegraphics[width=\linewidth]{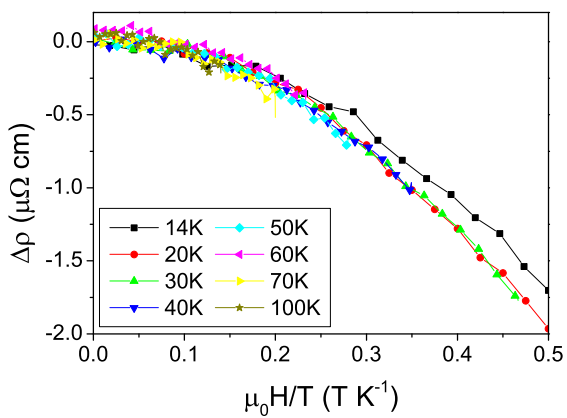}}
        \end{minipage}
    \caption{(a) MR of a high resistivity sample (\#2425), showing a (nearly) quadratic dependence. (b) The MR scales with H/T as expected for a spin flip scattering mechanism~\cite{Beal-Monod1968}.}
    \label{fig:Fig3}
\end{figure}

\par

Figure~\ref{fig:fig3a} shows a set of MR data obtained on a high resistivity sample (\#2425) above $T_{M}$. We have examined whether the MR data scales as a function of (H/T), as it does and is theoretically predicted in Kondo systems consisting of a metallic matrix and magnetic impurities~\cite{Monod1967,Beal-Monod1968}. Figure~\ref{fig:fig3b} shows that this scaling is well obeyed at temperatures ranging from about 15~K (outside the range of superconducting fluctuations) up to a temperature of about 90~K, somewhat below that where the negative MR cannot be detected anymore. The dependence on $(H/T)$ is nearly parabolic with a power law best-fit of 1.9. The MR of lower resistivity films also scales as a function of $(H/T)$ but the range of fields where it does so with an exponent close to 2 is limited.

\par

We have also checked the anisotropy of the MR, between magnetic field orientations parallel and perpendicular to the sample surface, while maintaining it perpendicular to the current in the sample. Both field configurations showed a negative MR being larger by about 30\% to 40\% in the parallel configuration. This anisotropy was found to be almost temperature independent.

\par

\begin{figure}
    \centering
    \begin{minipage}[t]{\linewidth}
    	\centering
        \subfigure[]{\label{fig:fig4a}\includegraphics[width=\linewidth]{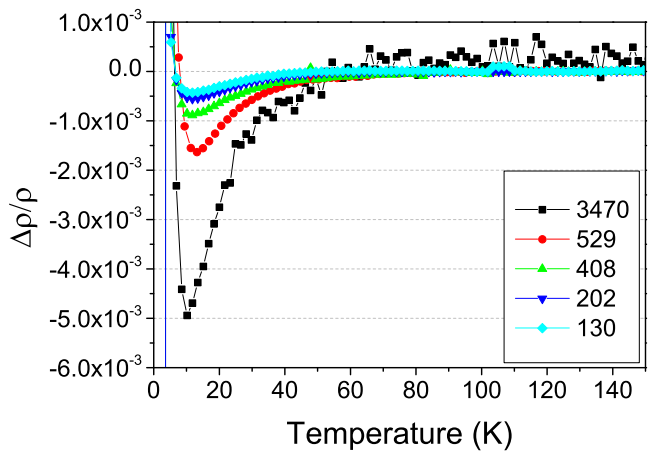}}
    \end{minipage}
    \begin{minipage}[b]{\linewidth}
    	\centering
        \subfigure[]{\label{fig:fig4b}\includegraphics[width=\linewidth]{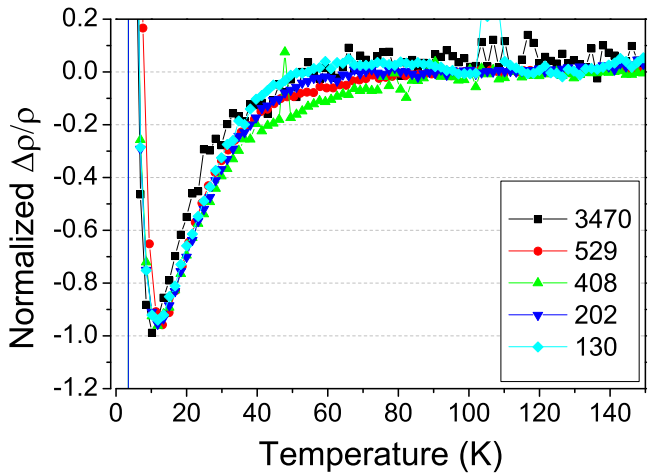}}
    \end{minipage}
    \caption{$\Delta\rho/\rho$ as a function of temperature for selected samples at a set field of 14~T. Figure (a) shows the negative MR enhancement with room temperature resistivity. Figure (b) shows that the temperature dependence is similar on all of our samples after scaling the raw data to the same maximum amplitude.}
    \label{fig:Fig4}
\end{figure}

We show in Fig.~\ref{fig:fig4a} how the MR amplitude varies as a function of temperature for a set field of 14~T, for a series of samples having different resistivities. The MR amplitude is seen to rise considerably with resistivity. But, as shown in Fig.~\ref{fig:fig4b}, all the data scale in the same way with temperature.

\par

A resistance minimum and logarithmic rise at low temperatures, and a negative MR, might also be due to weak electron localization~\cite{Deutscher1980}. However the observed $T^{-2}$ dependence of the negative MR rules out this interpretation. In weak localization (WL) theory the temperature dependence of the MR is given by that of the inelastic scattering time $\tau_{in}$. It is proportional to $\tau_{in}^{3/2}$ in 3D~\cite{KawabataA.1980}, with $\tau_{in}\propto T^{-p}$ where the value of $p$ depends on the scattering mechanism. For our range of temperatures, $T\geq 14K$, electron-phonon scattering is the dominant mechanism~\cite{Santhanam1984}, and $p=3$. This gives a temperature dependence of $T^{-9/2}$~\cite{KawabataA.1980}, which is inconsistent with our experimental result.

\par

Another important aspect of WL is the anisotropy of the MR. In 3D WL there should be no anisotropy. However, if one considers 2D WL~\cite{Deutscher1980}, an anisotropy should be observed. It depends on the relative strength of the spin-orbit and inelastic scattering times~\cite{Deutscher1982}. But there is no case where a negative MR is predicted for both magnetic field orientations, contrary to our results. Earlier MR measurements on granular Al, mostly performed at lower temperatures, also pointed out difficulties with a weak localization interpretation~\cite{Mui1984}. The negative curvature of R(T) at $T>T_{m}$ is another feature than cannot be explained by weak localization. Rather it reminds one of a similar effect seen in Kondo lattices~\cite{DeAlmeidaRibeiro2007,Hossain2002,Mizushima1999} and in underdoped cuprates~\cite{Takagi1992}.

\par

We thus turn back our attention to spin-flip scattering as a more likely origin of the observed negative magneto-resistance and Kondo-like behavior. It requires the interaction of conduction electrons with localized moments, as happens when certain impurities are in solution in a metallic matrix, for instance Fe in Cu. However, due to the high electronic density of states in Al, even Fe does not bare a magnetic moment in an Al matrix. We thus rule out the presence of magnetic impurity in Al as the origin of the magnetic moments interacting with conduction electrons.

\par

We see two other possible origins for magnetic moments interacting with conduction electrons in granular Al. May be the most obvious one is the presence of free spins at the Al/Al oxide interface, invoked by Sendelbach~\etal~\cite{Sendelbach2008} to explain the observed $1/f$ flux noise in SQUIDs. Interaction of these spins with conduction electrons has been postulated to explain the experimental results~\cite{Faoro2008}. A spin density $\sigma_{s}$ of $5\times10^{17}~m^{-2}$ has been calculated by Sendelbach~\etal, while Faoro and Ioffe suggest a value of $1\times10^{16}~m^{-2}$. The model of Faoro and Ioffe assumes that Al conduction electrons interact with interface magnetic moments. Therefore, a negative MR in a medium consisting of small Al grains and oxide interfaces is in agreement with their model.

\par

Another possible origin of localized magnetic moments would be the spins of unpaired electrons in small grains, on the condition that these grains have an electronic shell structure (otherwise the electronic wave functions of the single electrons are too similar to those of conduction electrons to produce a localized magnetic moment~\cite{Anderson1961}).

\par

The negative magneto-resistance for the case of dilute magnetic impurities in the small field limit ($\alpha =\frac{g{{\mu }_{B}}H}{{{k}_{B}}T}\ll 1$) is given by:
\begin{equation}
    {\Delta}{\rho} = -\frac{3\pi}{2\epsilon_{F}}\frac{m}{e^{2}\hbar}c\upsilon_{0}J^{2}\alpha^{2}u
    \label{eqn:DeltaRho}
\end{equation}
where $\epsilon_{F}$ is the Fermi energy, $c$ the magnetic impurities concentration, $\upsilon_{0}$ the atomic volume and $J$ the interaction parameter. $u$ is given as a function of the spin $S$ of the magnetic impurity, the Coulomb interaction $V$, $J$ and $\epsilon_{F}$ (Eq.~26 of ref.~\cite{Beal-Monod1968}). As remarked by B\'{e}al-Monod and Weiner the negative magneto-resistance is driven primarily by the progressive freezing out of spin-flip scattering when $\alpha$ is increased~\cite{Beal-Monod1968}. Although our granular films are quite different from a pure metallic matrix, we assume that this argument also applies to them. We retain the general idea that in the small field/high temperature limit the MR of a given sample should vary as $\alpha^{2}$. This immediately explains our central observation that the negative MR scales nearly as $T^{-2}$.

\par

We have used Eq.~\ref{eqn:DeltaRho}, in the form of $\Delta \rho =A{{\alpha }^{n}}$, in order to fit our MR data in the regime of $\alpha \lesssim 0.25\ll 1$. Here $A$ includes all the other parameters given in Eq.~\ref{eqn:DeltaRho}. In our case, the exponent of $\alpha $ varies from $1.75\pm 0.07$ in the low resistivity regime to $1.9\pm 0.02$ in the high resistivity regime for the given range of $\alpha $. We have compared our results of sample \#145 with the CuMn experimental MR data~\cite{Monod1967}, in order to estimate the magnetic impurity concentration, $c$. For the CuMn MR data $A\approx 4\times {{10}^{-3}}\mu \Omega cm$ while for sample \#145 we get $A\approx 178\times {{10}^{-3}}\mu \Omega cm$. This result suggests that the concentration $c$ in our sample is about 3400 ppm. Although $\epsilon_{F}$ and $\upsilon_{0}$ of Al are different than of Cu, and $J$ and $u$ are unknown for Al, we assume that they do not change this result by orders of magnitude. This concentration corresponds to about one spin per 2~nm grain. According to the surface spin densities given by Sendelbach~\etal~\cite{Sendelbach2008} and Faoro and Ioffe~\cite{Faoro2008} we would get between 0.1 and 6 spins per 2~nm grain, respectively.

\par

As seen from Fig.~\ref{fig:fig4a}, the amplitude of the negative MR increases rapidly as the MIT is approached. We can qualitatively understand it if the MIT is of the Mott type, as can be expected due to the large charging energy of the grains (about 0.1~eV for a 2~nm grain~\cite{Abeles1977}). This is because, as a Mott transition is approached, the Fermi energy and the electron mass are not anymore those of the parent metal. A narrow sub-band can form near the Fermi level, or in other terms by a decreasing Fermi energy which would contribute to an increase of the MR amplitude.

\par

In summary, the combined observations of a resistance minimum, a logarithmic resistance increase at low temperatures, a negative curvature of $\rho(T)$ at high temperatures, and the scaling of the negative magneto-resistance with (H/T) strongly suggest the presence of spin-flip scattering in granular Al films in the vicinity of the MIT.

\par

The presence of free spins can be attributed to surface effects at metal/oxide interfaces or to a volume effect by spins in the shell structure of the grains. The concentration of magnetic moments estimated from the measured magneto-resistance of films not too close to the MIT is compatible with values of the density of free spins at metal/oxide interfaces obtained from the ${1}/{f}\;$ flux noise seen in SQUIDS~\cite{Sendelbach2008}. Regardless of the origin of these localized magnetic moments, their measurable interaction with conduction electrons is a unique feature of nano-scale granular Al. In large Al grains of the order of 10 nm~\cite{Tse1978,*Tse1980} these moments are negligible and contribute no paramagnetic signal in the magnetic susceptibility.

\par

The interaction of conduction electrons with local magnetic moments should result in a decrease of the critical temperature, contrary to the increase seen in granular films Al films~\cite{Abeles1966,Cohen1968a,Deutscher1973a}. Our new findings showing the coexistence of magnetic properties in the normal state along with enhanced superconducting properties thus raise the question of the mechanism for superconductivity in these films. Because the change in sign of the magneto-resistance and the increase in the critical temperature occur simultaneously, one may wonder whether this mechanism does not involve an electron-electron interaction via spin fluctuations. Such a mechanism has just been proposed by Bodensiek~\etal~\cite{Bodensiek2013}, who have shown that near a Mott transition \textit{local} spin fluctuations can provide a mechanism for superconductivity with critical temperatures reaching several degrees Kelvin. This result was unexpected, as it was previously believed that only non local spin fluctuations could provide a mechanism for superconductivity. It fits well with our observations on granular films near the metal to insulator transition described here, with on average one spin per grain interacting with conduction electrons. The theoretical results described by Bodensiek~\etal~\cite{Bodensiek2013} were obtained on a Kondo lattice. However since the mechanism they describe is based on \textit{local} spin fluctuations it can be expected to apply to a disorder system such as ours.

\par

Further theoretical and experimental work is necessary to establish the exact origin of the spins interacting with conduction electrons in granular films, namely whether they originate from electronic level splitting due to the small grain size and a shell structure, or from the metal/oxide interfaces. In this respect a comparison between the behavior of granular films and that of non-granular thin films~\cite{Fortuin1997,*Santhanam1984,*Santhanam1987,*Bruynseraede1983} will be useful.

\par

In conclusion, we have demonstrated the co-existence of spins interacting with conduction electrons and an enhanced $T_c$. This co-existence strongly suggests an unconventional mechanism for superconductivity in granular films, which may well be that described by Bodensiek~\etal~\cite{Bodensiek2013}.

\begin{acknowledgments}
We want to thank Efrat Shimshoni for suggesting a possible scaling behavior of the negative MR and Ze'ev Lindenfeld for useful discussions. Conversations with Philippe Nozi\`{e}res are gratefully acknowledged. We are indebted to Enrique Gr\"{u}nbaum for a critical reading of the manuscript. This work was partially supported by EOARD award no. FA8655-10-1-3011.
\end{acknowledgments}

\bibliographystyle{apsrev4-1}
\bibliography{PRL2012_MRScaling}

\end{document}